\documentclass[usenatbib,usegraphicx]{mn2e}
\usepackage{aas_macros,bm,multirow}
\usepackage[usenames,dvipsnames]{color}
\usepackage[fleqn]{amsmath}
\usepackage[charter]{mathdesign}
\usepackage[hyperfootnotes, hyperfigures, pdftex]{hyperref}
\hypersetup{bookmarksnumbered, bookmarksopen, colorlinks, citecolor=blue, pdfduplex=DuplexFlipLongEdge}
\hypersetup{pdftitle=BH Spin from Realistic Temperature Profiles, pdfauthor=Akshay Kulkarni, pdfsubject=Astrophysics}
\usepackage{bookmark}   

\newcommand{\oneby}[1]{\frac{1}{#1}}                      
\newcommand{\e}[1]{\times 10^{#1}}                        

\newcommand{\dd}[2][]{\frac{d #1}{d #2}}                  

\newcommand{\fignar}[4][]{   
   \begin{figure}
   \centering
   \includegraphics[#1]{#2}
   \caption{#4}
   \label{#3}
   \end{figure}
}

\newcommand{\referee}[1]{#1}

\defcitealias{NovikovThorne73}{NT}
\defcitealias{PageThorne74}{PT}
\defcitealias{Shafee+08a}{a}
\defcitealias{Shafee+08b}{b}
\defcitealias{Steiner+10a}{a}
\defcitealias{Steiner+10b}{b}

\title[BH Spin from Realistic Temperature Profiles]{Measuring Black Hole Spin by the Continuum-Fitting Method: Effect of Deviations from the Novikov-Thorne Disc Model}

\author[A.~K. Kulkarni et al.]{
Akshay K. Kulkarni$^1$\thanks{E-mail:
   akulkarni@cfa.harvard.edu (AKK),
   rpenna@cfa.harvard.edu (RFP),
   rshcherbakov@cfa.harvard.edu (RS),
   jsteiner@cfa.harvard.edu (JFS),
   rnarayan@cfa.harvard.edu (RN),
   as@camk.edu.pl (AS),
   yzhu@cfa.harvard.edu (YZ),
   jem@head.cfa.harvard.edu (JEM),
   swd@cita.utoronto.ca (SWD),
   jmckinne@stanford.edu (JCM)},
Robert F. Penna$^1$\footnotemark[1],
Roman V. Shcherbakov$^1$\footnotemark[1],\newauthor
James F. Steiner$^1$\footnotemark[1],
Ramesh Narayan$^1$\footnotemark[1],
Aleksander S\k{a}dowski$^2$\footnotemark[1],
Yucong Zhu$^1$\footnotemark[1],\newauthor
Jeffrey E. McClintock$^1$\footnotemark[1],
Shane W. Davis$^3$\footnotemark[1],
Jonathan C. McKinney$^4$\footnotemark[1]
\\ $^1$Harvard-Smithsonian Center for Astrophysics, 60 Garden St, Cambridge, MA 02138\\
$^2$Nicolaus Copernicus Astronomical Center, Bartycka 18, PL 00-716, Warsaw, Poland\\
$^3$Canadian Institute for Theoretical Astrophysics. Toronto, ON M5S3H4, Canada \\
$^4$Department of Physics and Kavli Institute for Particle Astrophysics and Cosmology, Stanford University, Stanford, CA 94305-4060, USA
}

\begin{document}
\maketitle

\begin{abstract}
The X-ray spectra of accretion discs of eight stellar-mass black holes have been analyzed to date using the thermal continuum fitting method, and the spectral fits have been used to estimate the spin parameters of the black holes. However, the underlying model used in this method of estimating spin is the general relativistic thin-disc model of Novikov \& Thorne, which is only valid for razor-thin discs. We therefore expect errors in the measured values of spin due to inadequacies in the theoretical model. We investigate this issue by computing spectra of numerically calculated models of thin accretion discs around black holes, obtained via three-dimensional general relativistic magnetohydrodynamic (GRMHD) simulations.  We apply the continuum fitting method to these computed spectra to estimate the black hole spins and check how closely the values match the actual spin used in the GRMHD simulations. We find that the error in the dimensionless spin parameter is up to about 0.2 for a non-spinning black hole, depending on the inclination. For black holes with spins of 0.7, 0.9 and 0.98, the errors are up to about 0.1, 0.03 and 0.01 respectively. These errors are comparable to or smaller than those arising from current levels of observational uncertainty. Furthermore, we estimate that the GRMHD simulated discs from which these error estimates are obtained correspond to effective disc luminosities of about \referee{$0.4-0.7$} Eddington, and that the errors will be smaller for discs with luminosities of 0.3 Eddington or less, which are used in the continuum-fitting method.  We thus conclude that use of the Novikov-Thorne thin-disc model does not presently limit the accuracy of the continuum-fitting method of measuring black hole spin.
\end{abstract}



\section{Introduction}

Astrophysical black holes are described by just two parameters: their mass $M$ and angular momentum $J$, with the latter usually expressed in terms of the dimensionless spin parameter $a_* = cJ/GM^2$. While the mass $M$ is relatively straightforward to obtain using dynamical measurements, the spin parameter $a_*$ is less so. In accreting black holes, however, emission from the inner disc gives us a handle on the spin. According to the model developed by \citet[hereafter \citetalias{NovikovThorne73}]{NovikovThorne73}\footnote{This is the relativistic generalization of the standard thin-disc model of \citet{ShakuraSunyaev73}.} for a razor-thin accretion disc around a black hole, viscous evolution causes the accreting matter to move slowly inward along nearly \referee{Keplerian} orbits until reaching the radius of the innermost stable circular orbit (ISCO), after which the gas plunges into the black hole on a dynamical timescale. Thus, the inner edge of the viscous accretion disc is predicted to be very close to the ISCO. This link between the radius of the ISCO $r_{\rm ISCO}$ and the radius of the inner edge of the disc $r_{\rm in}$ is well supported by empirical evidence that the inner radius is constant in disc-dominated states of black hole binaries (\citealt{McClintock+07}; \citealt{Steiner+10a}\citetalias{Steiner+10a}; and references therein), and by recent GRMHD simulations of thin accretion discs (\citealt{Shafee+08a}\citetalias{Shafee+08a}; \citealt{Penna+10}; but see \citealt{Noble+09, Noble+10}). \referee{Therefore, measuring $r_{\rm in}$ gives one an estimate of $r_{\rm ISCO}$. Since $r_{\rm ISCO}/M$ is a monotonic function of $a_*$ \citep[e.g.,][]{ShapiroTeukolsky83}, we can then obtain the spin of the black hole. This is the main technique currently being used to estimate the spins of stellar-mass black holes in binary systems.

One of the ways of measuring $r_{\rm in}$ involves\footnote{\referee{Another method involves fitting the profile of the relativistically broadened iron line \citep[e.g.,][]{Fabian+89, Fabian+00, Laor91, ReynoldsNowak03, BrennemanReynolds06}.}} fitting the thermal X-ray continuum spectrum from the disc with the \citetalias{NovikovThorne73} model spectrum (e.g., \citealt{Zhang+97, Shafee+06, Davis+06, Gou+09, Gou+10, Steiner+09, Steiner+10a}\citetalias{Steiner+10a}) using models such as \textsc{KERRBB} \citep{Li+05} and \textsc{BHSPEC} \citep{DavisHubeny06} in the data-analysis package \textsc{XSPEC} \citep{Arnaud96}. From the fit, one obtains $r_{\rm in}$, or equivalently, $a_*$, if suitable estimates of the black hole mass, inclination and distance are available \citep[e.g., see][]{Gou+09}. Both \textsc{KERRBB} and \textsc{BHSPEC} assume that the structure of the disc and its emission properties are described accurately by the \citetalias{NovikovThorne73} model.}


It is therefore clear that a crucial issue in black hole spin estimation via the continuum-fitting method is the \citetalias{NovikovThorne73} model and its reliability. How much do real accretion discs with finite thickness differ from the \citetalias{NovikovThorne73} disc? This question was addressed by \citet{Paczynski00} and \citet{AfshordiPaczynski03}, who argued that deviations from the NT model decrease monotonically with decreasing disc thickness and that thin discs with dimensionless thickness $|h/r|\ll1$ are well described by the model, \referee{if the viscosity parameter $\alpha \ll 1$}. Their argument was confirmed by detailed calculations carried out by \citeauthor{Shafee+08b} (\citeyear{Shafee+08b}\citetalias{Shafee+08b}).  This still leaves open the question of whether {\it magnetized} discs might deviate substantially from the NT model even at small disc thicknesses \citep{Krolik99}. A number of recent studies of magnetized discs using three-dimensional general relativistic magnetohydrodynamic (3D GRMHD) simulations, including \citeauthor{Shafee+08a} (\citeyear{Shafee+08a}\citetalias{Shafee+08a}), \citet{Noble+09, Noble+10} and \citet{Penna+10}, have explored this question. These authors estimate that the luminosity and stress of the inner regions of simulated discs differ from the \citetalias{NovikovThorne73} model by factors ranging from a few percent \citep{Penna+10} to as much as 20 percent \citep{Noble+09}.  The question then is how much this departure affects measurements of black hole spin.

We investigate this question using a very straightforward approach: we start with a disc model obtained via the above-mentioned GRMHD simulations (principally models \referee{similar to those described in} Penna et al. 2010), compute the disc emission as a function of radius using a local blackbody approximation (assuming a constant spectral hardening factor), and use ray-tracing to compute the spectra. We then fit these spectra using \textsc{KERRBB} and compare the resulting spin estimate with the spin that was used in the GRMHD simulation. Our goal is similar to that of \citeauthor{Shafee+08b} (\citeyear{Shafee+08b}\citetalias{Shafee+08b}), who performed the same analysis for a purely hydrodynamical disc, and of \citet{LiYuanCao10}; the important difference between the latter work and ours is that we use disc models obtained from GRMHD simulations, although we do not explore the effect of \referee{a finite photospheric height} in detail.
Analogous work (though with a pseudo-Newtonian, not GRMHD, code) on the systematic errors in spin estimates obtained by fitting the broad iron emission lines from the inner accretion disc has been done by \citet{ReynoldsFabian08}.

We begin in \autoref{sec:method} with a description of our method, and calculate in \autoref{sec:results} the error in black hole spin estimates due to deviations of GRMHD discs from the NT model. We discuss in \autoref{sec:obs} the observational uncertainties in black hole spin determination and compare these with the errors arising from use of the NT model. We conclude in \autoref{sec:conc} with a summary. Some technical details are discussed in Appendices \ref{app:match}, \ref{app:trans} and \ref{app:grid}.


\section{Method}
\label{sec:method}

\subsection{Calculation of Disc Temperature and Velocity Profiles from GRMHD Simulations}
\label{sec:grmhd}

We work in Boyer-Lindquist (BL) coordinates $x^\alpha = (t, r, \theta, \phi)$. We use geometric units where the speed of light $c$, the gravitational constant $G$ and the Planck constant are set to unity, and measure all lengths and times in units of the black hole mass $M$.

\referee{For this project we reran} the three-dimensional general relativistic magnetohydrodynamic (3D GRMHD) simulations of thin discs described in \citet{Penna+10}, for four values of the black hole spin: $a_* = 0$, 0.7, 0.9, 0.98. For completeness, we briefly review the simulation setup here. The simulations solve the GRMHD equations for plasma around a rotating black hole using the code HARM \citep{Gammie+03} with numerous recent improvements, including 3D capabilities \citep{McKinney06, McKinneyBlandford09}. The gas is initially in a torus in hydrodynamic equilibrium surrounding the black hole \citep{DeVilliers+03, Gammie+03}. \referee{The spin axes of the torus and the black hole are aligned. The torus is} seeded with a magnetic field consisting of four poloidal loops arranged in the radial direction. We use a polytropic equation of state for the gas, $p \propto \rho^\gamma$, where $p$, $\rho$ and $\gamma$ are the pressure, density and adiabatic index respectively, and choose $\gamma = 4/3$, as appropriate for a radiation pressure dominated disc.

To keep the disc thin, we use a simple cooling prescription that drives the gas to its initial entropy on a dynamical timescale\footnote{One change from \citet{Penna+10} is that in the present work we cool all the gas, including the gas in the corona, whereas in most of their simulations, \citeauthor{Penna+10} cooled only the disc region of the flow. The present simulations are similar to the ``no-tapering model'' described in \S5.7 and Fig. 13 of their paper.}. The energy removed by the cooling prescription is \referee{assumed to be completely lost by the accretion flow}; it has no dynamical effect on the accreting gas (the energy lost to cooling is of course tracked and is later used to compute the disc luminosity profiles shown in \autoref{fig:lum}). The disc thickness is specified by the quantity $|h/r|$, where $h$ is the density scale height of the disc above the midplane, $|h| = \int\rho\,|z|\,dz / \int\rho\,dz$, and $r$ is the cylindrical radius. Our simulated \referee{thin} discs have \referee{$|h/r| = 0.05$, 0.04, 0.05 and 0.08 respectively for $a_*=0$, 0.7, 0.9 and 0.98}.  Following \citet{Penna+10}, we perform a temporal and azimuthal average over the steady-state portion of the simulation results to average over the fluctuations introduced by turbulence, since we are interested in the mean behaviour of the accretion flow. Finally, since our discs are geometrically thin, we perform a density-weighted average in the polar direction to obtain the vertically-integrated disc structure. This process of collapsing the simulated disc into the equatorial plane circumvents the difficulty of defining a photosphere for the disc and calculating emission profiles along it. A proper treatment would require a full radiative transfer calculation \referee{\citep[e.g.,][]{Davis+05, Sadowski+10}}, which is beyond the scope of this work.

At the end of this process, both components required to calculate the spectra are available: the radial profile of the fluid four-velocity $u^\mu(r)$ in the equatorial plane and, from the energy removed by the cooling prescription\footnote{\referee{\label{foot:bound} We only include the energy removed from the bound gas, since including all the gas results in an overestimate of the luminosity \citep{Penna+10}.}}, the profile of the emitted flux $F(r) \equiv dE/(rdrd\phi dt)$ (where $E$ is the energy emitted from one side of the disc as measured by an observer at infinity). When calculating the spectrum, the effect of electron scattering in the disc is taken into account indirectly using a colour correction (or spectral hardening) factor $f_{\rm col}$. For the results presented here, we choose for simplicity a fiducial value of $f_{\rm col} = 1.7$ \citep{ShimuraTakahara95}.
Detailed models of disc atmospheres by \citet{Davis+05} and \citet{DavisHubeny06} indicate that $f_{\rm col}$ can vary between 1.4 and 1.7, but this extra sophistication is not necessary for the simple tests described in the present paper.

The flux profiles obtained from the GRMHD simulations are only reliable within a radius inside which the accretion flow has reached steady state. Outside this radius, which we call the inflow equilibrium radius $r_{\rm ie}$ (see \citealt{Penna+10} for a definition), we extend the profiles using the analytical disc model of \citet{PageThorne74}. The procedure we use is described in \autoref{app:match}. Within a certain range, the exact choice of $r_{\rm ie}$ does not affect the results of the extrapolation, as we show in that appendix.

\autoref{fig:lum} compares the luminosity profiles $d (L/\dot M) / d(\ln r)$ that we obtain to those in the standard \citetalias{NovikovThorne73} disc model, for four values of the spin: $a_* = 0$, 0.7, 0.9, 0.98. Here, $\dot M$ is the accretion rate, and the luminosity $L \equiv 2 dE/dt = 2 \int F r dr d\phi = 4\pi \int F r dr$, so $d (L/\dot M) / d(\ln r) = 4\pi r^2 F(r) / \dot M$ (the extra factor of 2 is to account for emission from both sides of the disc).

The \citetalias{NovikovThorne73} model has no radiation from inside the ISCO, whereas the simulations show some emission from this region. In addition, the peak of the emission in the simulated discs is seen to shift inward relative to the \citetalias{NovikovThorne73} model. These effects are similar to those obtained by \citet{Sadowski09} for slim discs. As explained in that work, for large enough accretion rates ($\gtrsim 0.3$ Eddington), the accretion flow starts becoming radiatively inefficient at moderate radii ($r \sim 10M - 30M$); as a result, some of the heat generated by viscous dissipation at larger radii is advected inward and released at smaller radii. Another important effect is that discs with finite thickness have a non-vanishing stress at the ISCO (in contrast to the razor-thin discs which the NT model considers for which the stress is expected to vanish). This stress leads to additional viscous dissipation at radii $r \sim r_{\rm ISCO}$. In our model, the inward shift in the emission peak due to both of these effects mimics a decrease in $r_{\rm ISCO}$ (see \autoref{fig:lum}), i.e., an increase in the predicted black hole spin. As a result, fitting the GRMHD disc spectrum using the NT model leads to an overestimate of the black hole spin, as we shall see in \autoref{sec:results}.

\fignar{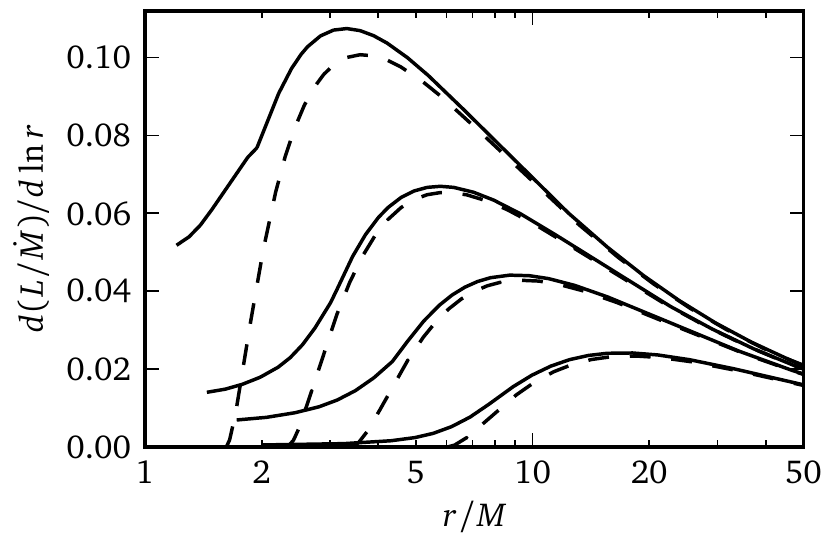}{fig:lum}{Luminosity profiles from the GRMHD simulations (solid lines) compared with those from the \citetalias{NovikovThorne73} model (dashed lines) for $a_*=0$, $0.7$, $0.9$ and $0.98$ (bottom to top). \referee{The disc thicknesses are $|h/r| = 0.05$, 0.04, 0.05 and 0.08 respectively for these runs.} The ISCO is located at the radius where the \citetalias{NovikovThorne73} disc luminosity goes to zero.
}

\subsection{Calculation of the Spectra}
To calculate the spectrum, we assume that the flux $F(r)$ is emitted in the form of colour-corrected blackbody radiation ($f_{\rm col} = 1.7$), either isotropically or with limb-darkening, as seen in the comoving frame of the fluid. We use a standard limb-darkening prescription (\autoref{eq:limbdark} below). We assume that after emission, the radiation propagates in vacuum.

Were the accretion disc non-relativistic, the calculation of the spectrum would be almost trivial \citep[see, e.g.,][]{FrankKingRaine02}; one would divide the disc into annuli, define an effective blackbody temperature $T_{\rm eff}(r) = [F(r)/\sigma]^{1/4}$ in each annulus, where $\sigma$ is the Stefan-Boltzmann constant, use the temperature and colour-correction factor to obtain the specific intensity $I_{\nu,\rm disc}(r)$ of the emitted radiation at the disc surface, and integrate it over the disc surface to obtain the observed spectrum.

Relativity introduces three complications: (1) The effective temperature has to be defined in the comoving frame of the fluid, and so we need to transform $F(r)$ from the BL frame into the comoving frame. (2) Redshift between the comoving frame and the observer's frame, both gravitational and due to Doppler boosting, has to be taken into account. Since the photon paths around a black hole are complicated, the direction in which a ray needs to be emitted in the comoving frame such that it reaches the observer is not known {\it a priori}, which is a problem for the redshift calculation. (3) One needs to know the emission direction to take limb darkening into account as well. Points (2) and (3) require integrating the geodesic equations to calculate the photon paths, which is usually referred to as ``ray-tracing.'' This approach has been applied extensively in the literature to a variety of problems, starting with \citet{CunninghamBardeen73} and \citet{Cunningham75} (see \citealt{DexterAgol09} and references therein). In particular,
\textsc{KERRBB} \citep{Li+05} uses this technique to compute thin-disc spectra.

\fignar{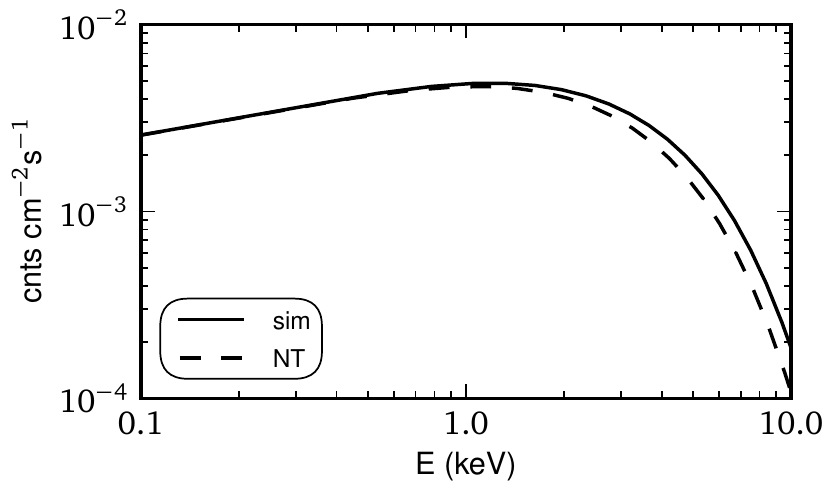}{fig:spec-1}{Spectra from the simulated (solid line) and \citetalias{NovikovThorne73} (dashed line) discs, for $a_*=0.9$ and $i = 75^\circ$.}

We perform ray-tracing numerically using the routines developed by \citet{ShcherbakovHuang11} and applied in \citet{Shcherbakov+10}. We choose a line of sight to the observer with an inclination angle of $i$ relative to the black hole spin axis. At a sufficiently large distance from the black hole ($r \sim 10^5$) we set up an image plane perpendicular to the line of sight and shoot rays from it parallel to the line of sight. We follow these rays until they hit the disc\footnote{This is more straightforward than shooting rays from the disc, since as mentioned earlier, the direction in which the rays need to be emitted from the disc such that they reach the observer is not known {\it a priori}. \referee{This approach was pioneered by \citet{Marck96}; see also \citet{Hameury+94}.}}, by directly integrating the (second-order) geodesic equations:
\begin{equation}
\dd[^2 x^\alpha]{\lambda^2} + \Gamma^\alpha_{\beta\gamma} \dd[x^\beta]{\lambda} \dd[x^\gamma]{\lambda} = 0,
\end{equation}
where $\lambda$ is an affine parameter along the geodesic, and $\Gamma^\alpha_{\beta\gamma}$ are the connection coefficients. The aim is to obtain the specific intensity $I_\nu$ of each ray, which can then be integrated over the image plane to obtain the observed flux:
\begin{equation}
   \label{eq:fobsdef}
   F_{\nu,\rm obs} = \oneby{D^2} \int I_\nu dA.
\end{equation}
Since $I_\nu / \nu^3$ is a Lorentz invariant, we can immediately relate the specific intensity $I_\nu$ in the image plane to the intensity $I_{\nu, \rm com}$ in the comoving frame of the fluid at the point of emission:
\begin{gather}
   I_\nu       = I_{\nu,\rm com} \left(\nu/\nu_{\rm com} \right)^3 \equiv I_{\nu,\rm com} \chi^3, \\
   \intertext{where $\chi \equiv \nu/\nu_{\rm com}$ is the redshift factor, and}
   I_{\nu,\rm com} = \frac{2 f_{\rm col}^{-4} \nu_{\rm com}^3}{\exp(\nu_{\rm com}/k_{\rm B} f_{\rm col} T_{\rm com}) - 1} \Upsilon .
\end{gather}
Here, $f_{\rm col}$ is the spectral hardening factor mentioned earlier, which we set to 1.7 in this work, and $\Upsilon$ is the limb-darkening factor \citep[see, e.g.,][]{Li+05}:
\begin{equation}
   \label{eq:limbdark}
   \Upsilon = \left\{
   \begin{aligned}
       & 1,                                         && \text{isotropic emission} \\
       & \frac{1}{2} + \frac{3}{4}\cos\theta_{\rm com}, && \text{limb-darkened emission}
   \end{aligned}
   \right.
\end{equation}
So finally we have
\begin{gather}
I_\nu = \frac{2 f_{\rm col}^{-4} \nu^3}{\exp(\nu/k_{\rm B} f_{\rm col} \chi T_{\rm com}) - 1} \Upsilon , \\
\intertext{or}
\label{eq:fobs}
F_{\nu,\rm obs} = \oneby{D^2} \int \frac{2 f_{\rm col}^{-4} \nu^3}{\exp(\nu/k_{\rm B} f_{\rm col} \chi T_{\rm com}) - 1} \Upsilon dA .
\end{gather}
Thus, to calculate the spectrum, we need the effective temperature in the comoving frame $T_{\rm com}$, the redshift factor $\chi$ and the angle $\theta_{\rm com}$ between the emitted ray and the disc normal in the comoving frame. The first of these is obtained by transforming the emitted flux $F(r)$, which is initially calculated in the Boyer-Lindquist frame, into the comoving frame, and the last two by transforming the ray four-momentum. We show the details in \autoref{app:trans}.

To calculate spectra using \autoref{eq:fobs}, we use the following fiducial parameters: black hole mass $M = 10M_\odot$, accretion rate $\dot M = 0.1 \dot M_{\rm Edd}$, and distance to the black hole $D = 10$ kpc. We choose a spectral energy range of 0.1 keV to 10 keV, divided into 1000 logarithmically spaced bins. \autoref{fig:spec-1} compares the spectra from the simulated and \citetalias{NovikovThorne73} discs for $a_*=0.9$ and $i=75^\circ$. The peak of the spectrum of the simulated disc is shifted to a slightly higher energy relative to the \citetalias{NovikovThorne73} spectrum, and the peak flux is also higher. This is precisely the effect that increasing the black hole spin would have on the \citetalias{NovikovThorne73} spectrum, and as we shall see in \autoref{sec:results}, fitting the simulated spectra leads one to overestimate the spin.

\autoref{fig:spec-i75} and \autoref{fig:spec-a9} show what happens to the difference between the simulated and \citetalias{NovikovThorne73} spectra when we vary the spin and the inclination respectively. The effect is visible at the high-energy end of the spectrum.
The effect of the inclination is very strong. This is because of the excess luminosity from the inner region of the simulated disc; this excess is more noticeable at higher inclinations due to beaming of the emitted radiation.


\fignar{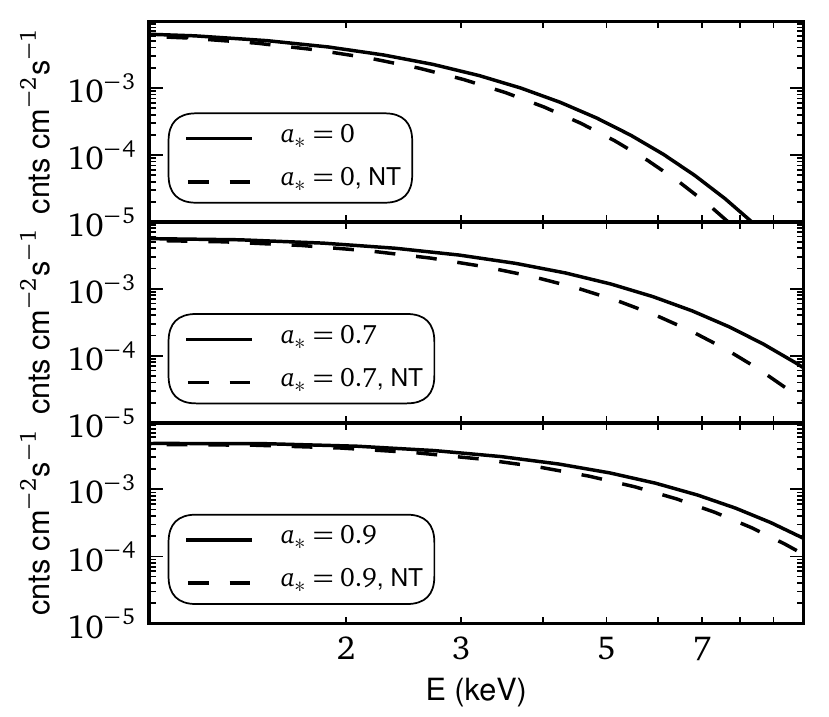}{fig:spec-i75}{High-energy portion of the spectra from the simulated (solid lines) and \citetalias{NovikovThorne73} (dashed lines) discs, for $i = 75^\circ$ and three values of the black hole spin: $a_*=0$, 0.7, 0.9.}

\fignar{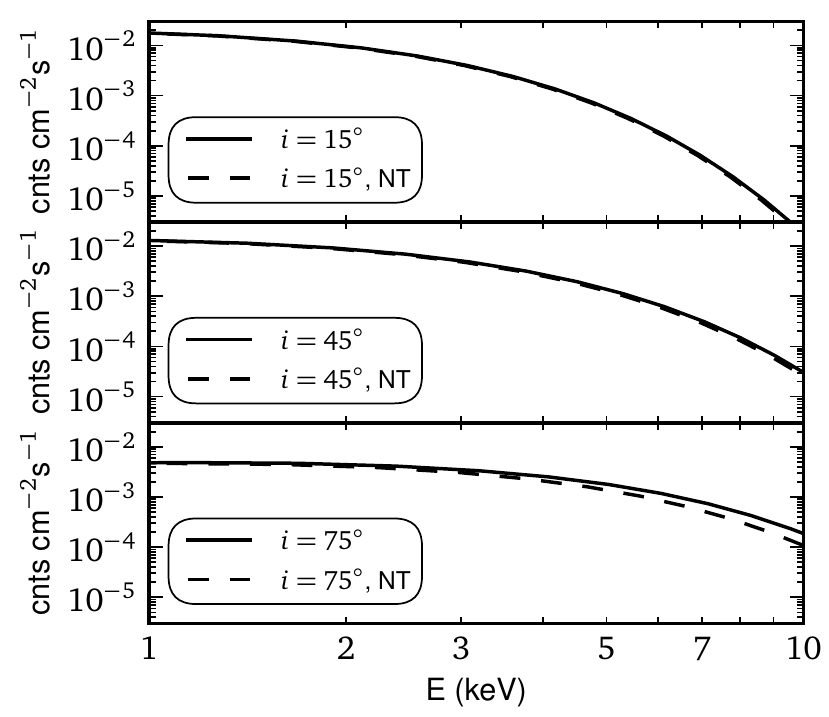}{fig:spec-a9}{High-energy portion of the spectra from the simulated (solid lines) and \citetalias{NovikovThorne73} (dashed lines) discs, for $a_*=0.9$ and three inclinations: $i=15^\circ, ~45^\circ, ~75^\circ$.}


\subsection{Tests}
We tested our code by comparing the spectra it produces for an \citetalias{NovikovThorne73} disc to those produced by \textsc{KERRBB} itself using the \verb+fakeit+ command in \textsc{XSPEC}, for the following range of parameters: black hole masses of $5, 10, 15~M_\odot$, spin parameters $a_* = 0, 0.7, 0.9$, observer inclinations $i = 15^\circ, 45^\circ, 75^\circ$, accretion rates $\dot{M}/\dot{M}_{\rm Edd} = 0.1, 0.2$ and distances $D = $10~kpc, 20~kpc, with and without using limb darkening. At a grid resolution of $N_b \times N_\beta = 100\times 100$ (see the description of our grid in \autoref{app:grid}), the spectra calculated with our code converge to the \textsc{KERRBB} spectra in all these cases. This confirms that the code is robust.


\section{Results}
\label{sec:results}
We use the following fiducial parameters for our spectra, as mentioned earlier: black hole mass $M = 10M_\odot$, accretion rate $\dot M = 0.1 \dot M_{\rm Edd}$, and distance to the black hole $D = 10$ kpc. The spectra are fitted using \textsc{KERRBB} (without using returning radiation, since we do not include it in our spectrum calculation) to obtain the black hole spin $a_*$ and accretion rate $\dot{M}$.

One more thing needs to be taken care of. Even though we average the GRMHD simulation results azimuthally and over time to remove the effects of turbulence and obtain a mean profile for the flux $F(r)$ and the gas four-velocity $u^\mu(r)$, there is still some stochastic variation in the spin estimates with time. Therefore, we divide the steady-state portion of each simulation into chunks of duration \referee{$\Delta t = 1000M$} (each simulation has \referee{4-5} such chunks), obtain a spin estimate from each chunk, and then quote the mean spin estimate and the error in the mean for each simulation\footnote{\referee{In addition, there are a couple of potential sources of systematic error: (i) our somewhat arbitrary choice of the matching radius used for extending the luminosity profiles beyond the inflow equilibrium radius (see \autoref{app:match}), and (ii) the fact that we restrict ourselves to the bound gas when calculating the luminosity profiles, as mentioned in \autoref{foot:bound}. We estimate the systematic error due to these two factors and, to be conservative, include them in quadrature in the error estimates that we quote in Tables \ref{tab:spins-main} and \ref{tab:mdot-main}.}}.

\begin{table}
\caption{Spin estimates obtained by fitting the simulated spectra (for \referee{limb-darkened} emission) with \textsc{KERRBB}, for a range of spins $a_*$ and observer inclination angles $i$. The model identified as ``1 loop'' corresponds to a GRMHD simulation that has one poloidal magnetic loop in its initial disc configuration; all the other models have four loops arranged radially.
}
\begin{centering}
\referee{
\begin{tabular}{lccc}
\hline
                 &   $a_*=0$           & $a_*=0$, 1 loop     &   $a_*=0.7$        \\
                 &   $|h/r|=0.05$      & $|h/r|=0.07$        &   $|h/r|=0.04$     \\
\hline
$i = 0^\circ$    &   0.08 $\pm$ 0.02   &   0.06 $\pm$ 0.01   &   0.71 $\pm$ 0.01  \\
$i = 15^\circ$   &   0.08 $\pm$ 0.02   &   0.06 $\pm$ 0.01   &   0.72 $\pm$ 0.01  \\
$i = 30^\circ$   &   0.09 $\pm$ 0.02   &   0.07 $\pm$ 0.02   &   0.72 $\pm$ 0.01  \\
$i = 45^\circ$   &   0.10 $\pm$ 0.02   &   0.09 $\pm$ 0.02   &   0.73 $\pm$ 0.01  \\
$i = 60^\circ$   &   0.11 $\pm$ 0.02   &   0.18 $\pm$ 0.01   &   0.76 $\pm$ 0.01  \\
$i = 75^\circ$   &   0.15 $\pm$ 0.04   &   0.37 $\pm$ 0.01   &   0.80 $\pm$ 0.02  \\
\hline
\end{tabular}
\begin{tabular}{lcc}
                 &   $a_*=0.9$           &   $a_*=0.98$         \\
                 &   $|h/r|=0.05$        &   $|h/r|=0.08$       \\
\hline
$i = 0^\circ$    &   0.905 $\pm$ 0.002   &   0.985 $\pm$ 0.001  \\
$i = 15^\circ$   &   0.906 $\pm$ 0.002   &   0.985 $\pm$ 0.001  \\
$i = 30^\circ$   &   0.907 $\pm$ 0.003   &   0.985 $\pm$ 0.001  \\
$i = 45^\circ$   &   0.908 $\pm$ 0.003   &   0.986 $\pm$ 0.001  \\
$i = 60^\circ$   &   0.914 $\pm$ 0.005   &   0.987 $\pm$ 0.001  \\
$i = 75^\circ$   &   0.929 $\pm$ 0.006   &   0.991 $\pm$ 0.001  \\
\hline
\end{tabular} \\
}
\end{centering}
\label{tab:spins-main}
\end{table}

\begin{table}
\caption{Absolute and fractional errors in the estimated radius of the innermost stable circular orbit, $r_{\rm ISCO}$, corresponding to the spin estimates in \autoref{tab:spins-main}.}
\begin{centering}
\referee{
\begin{tabular}{lcc}
\hline
                 &   $a_*=0$          &   $a_*=0.7$  \\
                 &   $r_{\rm ISCO}=6$ &   $r_{\rm ISCO}=3.39$ \\
\hline
$i = 0^\circ$    &   -0.26 (-4.3\%)   &   -0.07 (-2.0\%)  \\
$i = 15^\circ$   &   -0.27 (-4.5\%)   &   -0.07 (-2.2\%)  \\
$i = 30^\circ$   &   -0.29 (-4.9\%)   &   -0.10 (-2.9\%)  \\
$i = 45^\circ$   &   -0.33 (-5.4\%)   &   -0.15 (-4.5\%)  \\
$i = 60^\circ$   &   -0.37 (-6.2\%)   &   -0.28 (-8.2\%)  \\
$i = 75^\circ$   &   -0.49 (-8.2\%)   &   -0.51 (-15.1\%)  \\
\hline
                 &   $a_*=0.9$        &   $a_*=0.98$  \\
                 & $r_{\rm ISCO}=2.32$ & $r_{\rm ISCO}=1.61$ \\
\hline
$i = 0^\circ$    &   -0.04 (-1.6\%)   &   -0.07 (-4.1\%)  \\
$i = 15^\circ$   &   -0.04 (-1.7\%)   &   -0.07 (-4.2\%)  \\
$i = 30^\circ$   &   -0.04 (-1.9\%)   &   -0.07 (-4.5\%)  \\
$i = 45^\circ$   &   -0.06 (-2.5\%)   &   -0.08 (-5.2\%)  \\
$i = 60^\circ$   &   -0.10 (-4.1\%)   &   -0.11 (-6.9\%)  \\
$i = 75^\circ$   &   -0.21 (-9.0\%)   &   -0.18 (-11.0\%)  \\
\hline
\end{tabular} \\
}
\end{centering}
\label{tab:deltar-main}
\end{table}

The results for limb-darkened emission\footnote{We also looked at spectra generated using isotropic emission. To fit these spectra we turned off the limb-darkening flag of \textsc{KERRBB}. The resulting spin estimates are very similar to those obtained using limb-darkened emission, so we do not show them here.} are shown in columns 1, 3, 4 and 5 of \autoref{tab:spins-main}. As expected, the fitted values are different than the ones used in the GRMHD simulations. The differences are largest at low spins. It is easy to understand why the difference is not constant; the dependence of the disc temperature profile (which determines the shape of the spectrum) on the spin is highly nonlinear. In particular, the position of the spectral peak in the \citetalias{NovikovThorne73} model strongly depends on the radius of the ISCO ($r_{\rm ISCO}$), to the extent that one can think of \textsc{KERRBB} as fitting for $r_{\rm ISCO}$ instead of $a_*$. There is a one-to-one relationship between $a_*$ and $r_{\rm ISCO}$ \citep[see., e.g.,][]{ShapiroTeukolsky83}, \referee{shown in \autoref{fig:a-risco}. At high spins, $r_{\rm ISCO}$ varies very rapidly as a function of $a_*$; conversely, $a_*$ varies relatively slowly as a function of $r_{\rm ISCO}$. Thus, a given fractional error in $r_{\rm ISCO}$ translates into a much smaller error in $a_*$ at higher spins than at smaller ones. This is illustrated by the two gray bands in \autoref{fig:a-risco}. Each band represents a range of $\pm10\%$ in $r_{\rm ISCO}$, around $r_{\rm ISCO} = 6$ (upper band, corresponding to $a_*=0$) and $r_{\rm ISCO} = 1.61$ (lower band, corresponding to $a_*=0.98$). The corresponding range in $a_*$ is $\pm0.2$ at $a_*=0$, but only $\pm0.01$ at $a_*=0.98$.}

For reference, we show the errors in the estimated $r_{\rm ISCO}$ in \autoref{tab:deltar-main} (the exact values of $r_{\rm ISCO}$ for $a_*=0$, 0.7, 0.9 and 0.98 are 6, 3.39, 2.32 and 1.61), and the fitted accretion rates in \autoref{tab:mdot-main} (the fiducial rate of $0.1\dot M_{\rm Edd}$ used here corresponds to $(2.45, 1.35, 0.898, 0.5982) \e{18}~{\rm g\,s^{-1}}$ for the four spins, assuming that the accretion efficiencies are given by their \citetalias{NovikovThorne73} values). It is interesting to note that although the errors in the fitted spins are relatively large, the errors in the accretion rates are only a few percent.

\fignar{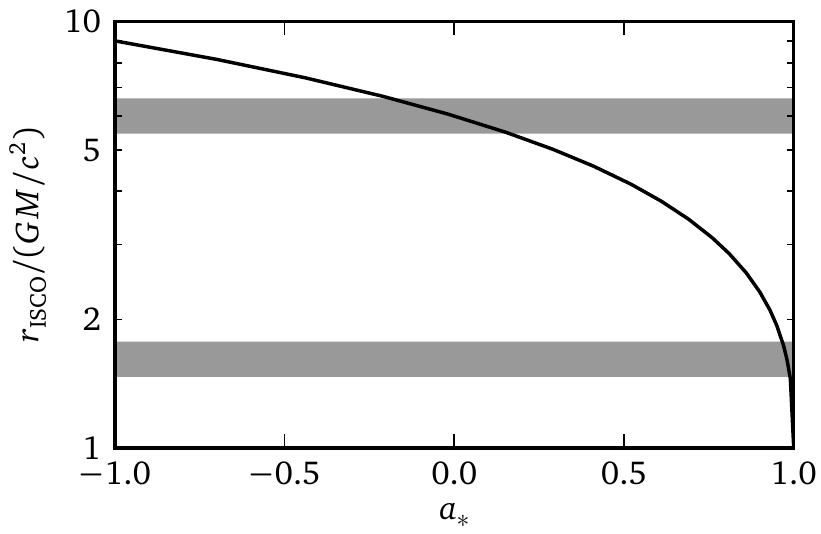}{fig:a-risco}{Relation between the black hole spin parameter $a_*$ and the radius of the innermost stable circular orbit $r_{\rm ISCO}$.}

The other important effect is that of the observer inclination: at high inclination, the error in the spin estimate is larger. This is because the difference between the disc temperature profiles in the \citetalias{NovikovThorne73} model and the simulations is significant only in the inner disc. At low inclination angles, the combined effect of gravitational redshift and beaming of the radiation (the latter of which concentrates the radiation close to the equatorial plane) results in this difference not being noticeable in the spectrum. At high inclination angles, on the contrary, beaming enhances the difference, causing the error in the fitted spin to increase.

Changing the black hole mass, accretion rate or distance only changes the overall scaling of the spectrum; therefore, there is no effect on the shape of the spectrum or the spin estimates. We should note, however, that since the GRMHD simulations use dimensionless quantities, the gas mass scale in the simulations is arbitrary. Therefore, we have the ability to choose any accretion rate for a given disc thickness. For real discs, this is certainly not the case. The relation between the disc thickness and the luminosity
is discussed in more detail in \autoref{sec:slimdisk}. We show there that the disc thicknesses used in our simulations (\referee{$|h/r|=0.05$, 0.04, 0.05 and 0.08} for $a_*=0$, 0.7, 0.9 and 0.98 respectively) correspond to \referee{$L/L_{\rm Edd} = 0.5$, 0.4, 0.5 and 0.7 respectively}; therefore, strictly speaking, our estimates of the errors in the spin determination are only applicable for these luminosities.

\begin{table}
\caption{Fitted accretion rates for the cases in \autoref{tab:spins-main}, in units of $10^{18}~{\rm g\,s^{-1}}$. The input values in the GRMHD simulations are denoted by $\dot M_{\rm input}$.
}
\begin{centering}
\referee{
\begin{tabular}{lcc}
\hline
                 &   $a_*=0$             &   $a_*=0.7$            \\
                 & $\dot M_{\rm input} = 2.45$ & $\dot M_{\rm input}=1.35$ \\
\hline
$i = 0^\circ$    &   2.46 $\pm$ 0.01     &   1.37 $\pm$ 0.01      \\
$i = 15^\circ$   &   2.46 $\pm$ 0.01     &   1.37 $\pm$ 0.01      \\
$i = 30^\circ$   &   2.46 $\pm$ 0.01     &   1.37 $\pm$ 0.01      \\
$i = 45^\circ$   &   2.47 $\pm$ 0.01     &   1.36 $\pm$ 0.01      \\
$i = 60^\circ$   &   2.48 $\pm$ 0.01     &   1.34 $\pm$ 0.02      \\
$i = 75^\circ$   &   2.49 $\pm$ 0.04     &   1.30 $\pm$ 0.04      \\
\hline
                 &   $a_*=0.9$           &   $a_*=0.98$           \\
                 & $\dot M_{\rm input}=0.898$ & $\dot M_{\rm input}=0.5982$ \\
\hline
$i = 0^\circ$    &   0.901 $\pm$ 0.001   &   0.5983 $\pm$ 0.0003  \\
$i = 15^\circ$   &   0.901 $\pm$ 0.001   &   0.5982 $\pm$ 0.0004  \\
$i = 30^\circ$   &   0.902 $\pm$ 0.001   &   0.5983 $\pm$ 0.0005  \\
$i = 45^\circ$   &   0.903 $\pm$ 0.001   &   0.5985 $\pm$ 0.0005  \\
$i = 60^\circ$   &   0.904 $\pm$ 0.003   &   0.5980 $\pm$ 0.0008  \\
$i = 75^\circ$   &   0.892 $\pm$ 0.007   &   0.594 $\pm$ 0.001  \\
\hline
\end{tabular} \\
}
\end{centering}
\label{tab:mdot-main}
\end{table}

We carried out a test run at $a_*=0$ with a different initial magnetic field configuration that has one poloidal loop instead of four as in our other runs. This model is closer in spirit to the simulations run by \citet{Noble+09, Noble+10}. We find that the one-loop model gives hotter spectra and a larger error in the derived value of the spin \referee{at large inclination angles} (compare the first two columns of \autoref{tab:spins-main}).
This agrees with the results described by \citet{Noble+10} and \citet{Penna+10}, who investigated the behavior of other diagnostics such as the angular momentum and shear stress and showed that GRMHD discs calculated from single-loop initial conditions generally deviate more strongly from the NT model compared to discs obtained from multi-loop initial conditions. \citet{Penna+10} argued that the multi-loop case is more natural since it better mimics disc turbulence, whereas the one-loop case might introduce an artificial long-range radial coherence in the solution.

The errors in the spin estimates could be due to a number of reasons: (1) The disc emissivity profile outside the ISCO is different in the simulations compared to the NT  model, as \autoref{fig:lum} shows; (2) The simulations have some radiation coming from the plunging region inside the ISCO; and (3) Even outside the plunging region, the radial component of the gas four-velocity is not negligible. To find out which of these is the dominant effect, we calculated some spectra from the simulated discs by excluding the region inside the ISCO and setting the gas velocity outside the ISCO to its \citetalias{NovikovThorne73} value. Any residual differences in the spin estimates would solely be due to (1).

The results are shown in \autoref{tab:spins-uNT-exclude}. We see that spin estimates obtained from the GRMHD simulations are still significantly different from the true values. This shows that the dominant reason for the errors in the spin estimates is the fact that, even outside the ISCO, the disc emissivity profile in the simulations is different from the \citetalias{NovikovThorne73} profile; more specifically, that the peak of the profile is shifted to smaller radii, as mentioned in \autoref{sec:method}. \referee{We should note, however, that for discs thicker than those considered in this work by about a factor of 2 or more, the effect of the plunging region is important. This follows from the finding of \citet{Penna+10} that deviations of the GRMHD simulations from the \citetalias{NovikovThorne73} model increase with increasing disc thickness.}


\subsection{Effective Accretion Rates of the GRMHD Models}
\label{sec:slimdisk}

The GRMHD models that we have used in this study make use of dimensionless quantities and do not include detailed radiation transfer. Hence there is no direct way of estimating the physical mass accretion rate (${\rm g\,s^{-1}}$) or the true radiative luminosity (${\rm erg\,s^{-1}}$) of the models. To estimate these quantities we use an indirect method, in which we compare the vertical thicknesses of the simulated discs against physical disc models that do include radiation transfer and radiation pressure and solve for the vertical disc structure.

We use two models for this comparison. One is a semi-analytical model of a slim disc \citep{Sadowski+10} which goes beyond the NT model by including the effect of energy advection in the radial equations. At each radius $r$, the model solves the condition of vertical hydrostatic equilibrium and includes radiative transfer approximately. The other model \citep{Davis+05} assumes the NT model for the radial structure but carries out a careful and detailed computation of radiation transfer, including non-LTE effects, at each $r$.  This model is identical to the \textsc{XSPEC} model \textsc{BHSPEC} \citep{DavisHubeny06}.  Each of these models treats some part of the physics very well, but neither has all the ingredients one would like to include, viz., advection, full radiative transfer, magnetic fields, deviations from hydrostatic equilibrium, self-irradiation, etc.

\begin{table}
\caption{Spin estimates from spectra obtained by excluding the plunging region and setting the four-velocity in the disc to its \citetalias{NovikovThorne73} value (second column) compared with the original spin estimates from \autoref{tab:spins-main} (third column).}
\begin{centering}
\referee{
\begin{tabular}{ccc}
\hline
$a_* = 0.9$, $i = 75^\circ$  & 0.92 & 0.93\\
$a_* = 0.9$, $i = 45^\circ$ & 0.91 & 0.91 \\
$a_* = 0$, $i = 75^\circ$ & 0.13 & 0.15 \\
\hline
\end{tabular} \\
}
\end{centering}
\label{tab:spins-uNT-exclude}
\end{table}

\autoref{fig:hr} shows the disc thickness as a function of $r$ for $a_*=0$, as predicted by the slim-disc model (solid lines, corresponding to $L/L_{\rm Edd} = 0.3$, 0.4, 0.5, bottom to top) and \textsc{BHSPEC} (dashed lines, same set of luminosities). These curves should be compared with the disc thickness in the GRMHD simulation (dotted lines). The top panel shows the density scale height $|h| = \int\rho\,|z|\,dz / \int\rho\,dz$, while the bottom panel shows the rms height $h_{\rm rms} = (\int\rho\,z^2\,dz / \int\rho\,dz)^{1/2}$.

At radii close to the ISCO, the slim disc and \textsc{BHSPEC} models indicate that the disc thickness plunges to small values whereas the GRMHD simulation shows a much smaller decrease. We believe there are at least three reasons for this discrepancy: \referee{(i)~the GRMHD simulations cool the gas by forcing it towards a constant entropy, which may not be justified in the plunging region;} (ii) the simulated GRMHD disc includes magnetic fields whose pressure provides additional support in the vertical direction whereas the other two models do not; and (iii) the simulated disc begins to deviate from hydrostatic equilibrium as the radial velocity becomes large near the ISCO and the gas has less time to reach equilibrium, whereas the other models hardwire the condition of hydrostatic equilibrium at all radii. \referee{We estimate that the last two are only important well inside the ISCO, however, while the discrepancy sets in already at larger radii.} These are interesting issues which we hope to explore in the future.  For the purposes of this section, we simply ignore the region of the simulation near the ISCO.

For the comparisons described here, we select a radius of $r=12M=2 r_{\rm ISCO}$, which is well outside the ISCO, and determine the luminosities at which the slim disc and \textsc{BHSPEC} models give the same disc thickness as we obtain in the simulated disc.  We see from \autoref{fig:hr} that the thickness measure $|h/r| \sim 0.05$ in the simulated GRMHD disc corresponds to $L/L_{\rm Edd} \sim 0.5$ according to the slim-disc model and $\sim 0.4$ according to \textsc{BHSPEC}. A comparison of the thickness measure $h_{\rm rms}/r$ gives slightly larger values of $L/L_{\rm Edd}$. Similar analysis (again at $r=2 r_{\rm ISCO}$) for $a_*=$ 0.7, 0.9 and 0.98 shows that \referee{$L/L_{\rm Edd} \sim$ 0.4, 0.5 and 0.7} respectively according to the slim-disc model, and \referee{0.4}, 0.4 and \referee{0.6} respectively according to \textsc{BHSPEC}.

\fignar{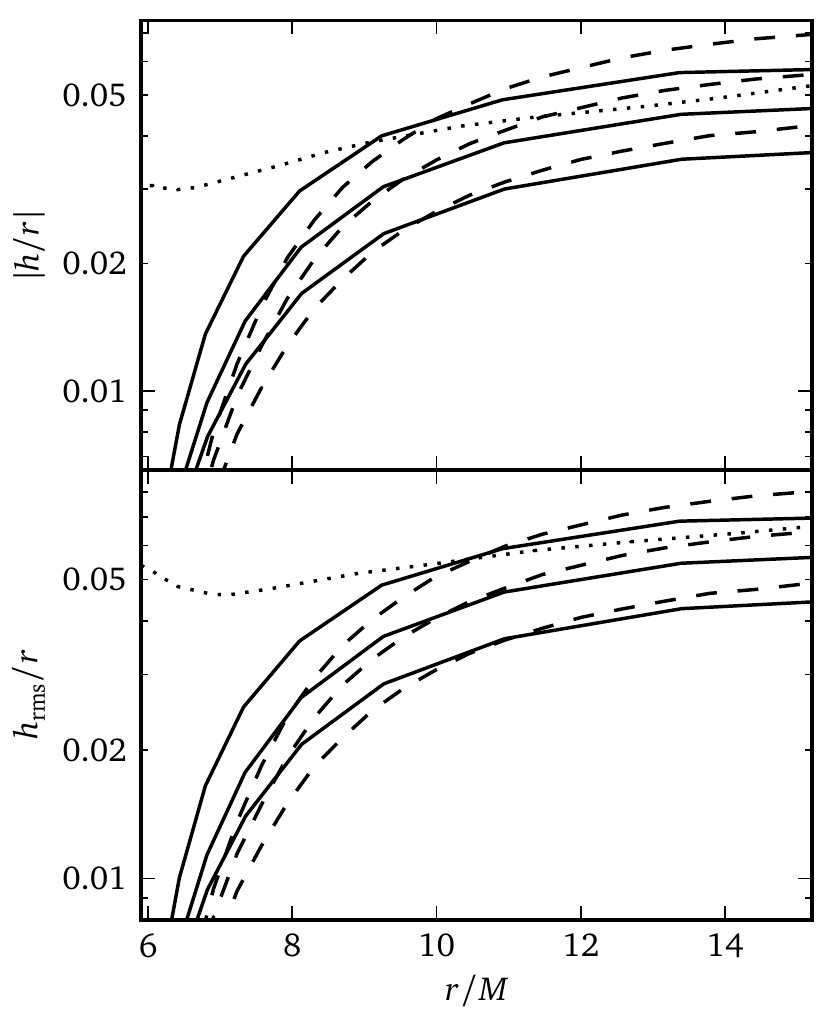}{fig:hr}{Profiles of $|h/r|$ for $a_*=0$ from the slim-disc model of \citet{Sadowski+10} (solid lines) and the detailed radiative transfer model \textsc{BHSPEC} of \citet{DavisHubeny06} (dashed lines) for luminosities $L/L_{\rm Edd} =$ 0.3, 0.4, 0.5 (bottom to top in each panel), compared with the profile obtained from the GRMHD simulation (dotted lines). The top panel shows the density scale height $|h| = \int\rho\,|z|\,dz / \int\rho\,dz$, while the bottom panel shows the rms height $h_{\rm rms} = (\int\rho\,z^2\,dz / \int\rho\,dz)^{1/2}$.}

The above luminosities are much higher than the typical luminosities ($L/L_{\rm Edd}<0.3$) at which black hole binary spectra are analyzed for spin determination using the continuum-fitting method (from \autoref{fig:hr}, this corresponds to $|h/r| \lesssim 0.03$). Therefore, the errors in the spin determination quoted in \autoref{sec:results} are not directly applicable to observations. At lower luminosities, the accretion disc would be thinner, as \autoref{fig:hr} shows. Due to computational resource requirements, it is not currently possible for us to perform GRMHD simulations for discs thinner than those presented here. What we can do instead is to scale the results from the simulations to more realistic thinner discs at lower luminosities. \autoref{tab:hr2} compares the spins obtained from two GRMHD simulations corresponding to $a_*=0$, one with \referee{$|h/r|=0.05$} (which we have focused on so far) and another with \referee{$|h/r| = 0.09$}. We see that the error in the spin estimate is much larger for the latter model.
Thus, it is clear that at the lower luminosities that are interesting from an observational point of view, the errors in the spin estimates would be \referee{significantly} smaller than those in \autoref{tab:spins-main}.

As an aside, we note that \citet{McClintock+06} calculated the height of the disc photosphere as a function of radius for an \citetalias{NovikovThorne73} disc (see Figure 17 in their paper). The disc heights they quote are larger than the density scale heights shown in \autoref{fig:hr} by roughly a factor of 2--3. We have calculated the location of the photosphere corresponding to the slim disc and \textsc{BHSPEC} models, and find that they agree fairly well with the values obtained by \citet{McClintock+06}.

\begin{table}
\caption{Spin estimates for $a_*=0$ from GRMHD simulations of a thicker disc \referee{($|h/r| = 0.09$)} compared with those for the thinner \referee{($|h/r| = 0.05$)} disc shown in \autoref{tab:spins-main}.}
\begin{centering}
\referee{
\begin{tabular}{ccccc}
\hline
$|h/r|$ & $L/L_{\rm Edd}$ & $i = 15^\circ$  & $i = 45^\circ$ & $i = 75^\circ$ \\
\hline
0.05 & 0.5 & 0.08 & 0.10 & 0.15 \\
0.09 & 0.9 & 0.08 & 0.19 & 0.50 \\
\hline
\end{tabular} \\
}
\end{centering}
\label{tab:hr2}
\end{table}


\begin{table}
\caption{Spin estimates from spectra calculated using a finite \referee{photospheric height} for the disc ($|h_{\rm phot}/r| = 0.2$) for both \citetalias{NovikovThorne73} and simulated disc temperature profiles, at selected values of spin and inclination.}
\begin{centering}
\referee{
\begin{tabular}{lccc}
\hline
                        & \citetalias{NovikovThorne73} disc & simulated & simulated disc \\
                        &                                   & disc      & (original estimates \\
                        &                                   &           & from \autoref{tab:spins-main}) \\
\hline
$a_* = 0.9$, $i = 75^\circ$                      & 0.88  & 0.90 & 0.93 \\
$a_* = 0.9$, $i = 45^\circ$                      & 0.89  & 0.90 & 0.91 \\
$a_* = 0$, $i = 75^\circ$                        & -0.05 & 0.09 & 0.15 \\
\hline
\end{tabular} \\
}
\end{centering}
\label{tab:spins-thick}
\end{table}

\begin{table*}
\caption{Data for four black holes: The entries are respectively the number of spectra analyzed; inclination angle; spin parameter; the approximate/symmetrized absolute and fractional errors in $r_{\rm ISCO}$; and the Eddington-scaled luminosity.}
\begin{centering}
\begin{tabular}{lcccccl}
\hline
Black Hole& No.& $i$ (deg)& $a_*$& $\Delta{r}$($\Delta{r}/r$)& $L/L_{\rm Edd}$& Reference \\
\hline
A0620--00&      1&       $51.0\pm0.9$&  $0.12\pm0.19$&          $\pm0.65$(11.5\%)& 0.11&       \citet{Gou+10} \\
XTE J1550--564& 45$^a$&  $74.7\pm3.8$&  $0.34_{-0.28}^{+0.20}$& $\pm0.86$(17.9\%)& 0.05--0.30& \citeauthor{Steiner+10b} (\citeyear{Steiner+10b}\citetalias{Steiner+10b}) \\
M33 X-7&        15&      $74.6\pm1.0$&  $0.84\pm0.05$&          $\pm0.29$(10.7\%)& 0.07--0.11& \citet{Liu+08, Liu+10} \\
LMC X-1&        18&      $36.4\pm2.0$&  $0.92_{-0.07}^{+0.05}$& $\pm0.43$(20.5\%)& 0.15--0.17& \citet{Gou+09} \\
\hline
\end{tabular} \\
$^a$ Typical value: Number of spectra analyzed varies depending on details of data selection (see \citeauthor{Steiner+10b} \citeyear{Steiner+10b}\citetalias{Steiner+10b}).
\end{centering}
\label{tab:obs}
\end{table*}

\subsection{Effect of a Finite Photospheric Height}
\label{sec:finite-thick}

So far we have been calculating spectra using equatorial profiles of the emitted flux $F(r)$ and the fluid four velocity $u^\mu(r)$, which are obtained, as mentioned in \autoref{sec:method}, by vertically integrating the GRMHD simulated disc structure. This integration effectively collapses the disc into the equatorial plane, which is where the disc emission is assumed to originate from. The errors in the spin estimates quoted above have therefore been purely due to the departure of the equatorial flux and fluid velocity from their \citetalias{NovikovThorne73} values. In reality, however, the observed disc emission comes from the photosphere, which is at a finite height above the equatorial plane. The effect of this on the spin estimates needs to be checked.

The method we use for doing this is very crude; our only goal here is to find out whether or not this effect could be important. The slim disc and \textsc{BHSPEC} models mentioned above, and the calculations of \citet{McClintock+06}, show that the photosphere height is about 2--3 scale heights. Since we want to see how large the effect of off-midplane emission could possibly be, we choose a larger photosphere height: $h_{\rm phot}/r \sim 4|h/r|$, where $|h/r|$ is the disc half-thickness measured at one scale height above the disc midplane. We then repeat our ray-tracing computation, but following the geodesics until they hit the photosphere instead of the equatorial plane (the photosphere as defined above corresponds to $\theta = \pi/2 - \delta$, where $\delta = \tan^{-1}(h_{\rm phot}/r)$). Finally, we assume that the flux and gas four-velocity profiles at the point of emission are given by their equatorial values $F(r)$ and $u^\mu(r)$, enabling us to calculate the spectra. \referee{\autoref{tab:spins-thick} indicates that the effect on the estimates of $a_*$ may be significant at high spins. However, it is encouraging to note that the errors in the spin estimates decrease when we use a finite photospheric height.}

We should \referee{also note} that \citet{LiYuanCao10} find a much larger effect on the spin estimates when they take the \referee{effect of a finite photospheric height} into account. We believe this is due to the fact that the $|h/r|$ profile in their analysis drops relatively sharply at small radii, like the profiles from the slim disc model and \textsc{BHSPEC} shown in \autoref{fig:hr}. This disc geometry leads to self-shadowing of the disc. For our analysis in this subsection, on the other hand, we have chosen a constant $|h/r|$. Even if we were to use the $|h/r|$ profile from the GRMHD simulations, we would not expect to see significant self-shadowing, since the $|h/r|$ profile in the simulated model is nearly constant with radius (\autoref{fig:hr}).


\section{A Comparison of Observational and Model-Dependent Errors}
\label{sec:obs}

The obvious question at this point is how big the errors in the spin estimates listed in \autoref{tab:spins-main} are compared with the observational uncertainties in spin determination. We address this question in this section.  The spins of eight stellar-mass black holes have been measured so far using the continuum-fitting method.  The observational error estimates for the first four (see \citealt{Shafee+06} for GRO J1655--40 and 4U 1543--47, \citealt{Davis+06} for LMC X-3, and \citealt{McClintock+06} for GRS 1915+105) are very approximate, and we disregard these results here.  In more recent work, the principal sources of observational errors, as well as the uncertainties in the key model parameters (e.g., the viscosity parameter $\alpha$), have been treated in detail. Moreover, in a recent paper on XTE J1550--564, \citet{Steiner+10b} have exhaustively explored many additional sources of error (see their Table~3 and Appendix A). The upshot of the work to date is that in every case the uncertainty in $a_*$ is completely dominated by the errors in three key dynamical parameters that are input when fitting the X-ray spectral data \citep{McClintock+06}.  These parameters are the distance $D$, the black hole mass $M$, and the inclination of the inner disc $i$ (which is assumed to be aligned with the orbital angular momentum vector of the binary; \citealt{Li+09}).  In order to determine the error in $a_*$ due to the combined uncertainties in $D$, $M$ and $i$, Monte Carlo simulations are performed assuming that these parameters are normally and independently distributed \citep[e.g.,][]{Gou+09}.

\autoref{tab:obs} gives selected observational data for four black holes (all of these have been subjected to the rigorous error analysis described above): the inclination angle, which has an important effect on the model results (Tables \ref{tab:spins-main}--\ref{tab:spins-thick}); the spin parameter; the absolute and fractional errors in $r_{\rm ISCO}$ (compare \autoref{tab:deltar-main}); and the luminosity.  All errors are quoted at the 68\% level of confidence.  Note that the values of $a_*$ range widely from $\sim0$ to $\sim0.9$.  As a rough characterization, the uncertainties in the values of $a_*$ are $\Delta{a_*} \sim \pm0.05$ for the rapidly spinning pair of black holes and $\Delta{a_*} \sim \pm0.2$ for the slowly spinning pair.  The corresponding fractional errors in $r_{\rm ISCO}$ range from approximately 10\% to 20\%.  Comparing the fractional errors in $r_{\rm ISCO}$ in \autoref{tab:obs} with the closest counterpart results in \autoref{tab:deltar-main} (i.e., closest matches for $i$ and $a_*$), we find that the error in the NT model is in all cases less than the observational error: \referee{A0620--00, 5.4\% vs.\ 11.5\%; XTE J1550--564, 8.2\% vs.\ 17.9\%; M33 X-7, 9.0\% vs.\ 10.7\%; and LMC X-1, 1.9\% vs.\ 20.5\%.}

Furthermore, the estimates of the modeling error due to deviations from the NT model obtained in this paper are very likely overestimates because the GRMHD simulation results necessarily correspond to relatively luminous discs: \referee{$L/L_{\rm Edd} = 0.4-0.7$}, whereas the observed luminosities are typically only $L/L_{\rm Edd} \sim 0.15$ (Table~7) and are strictly limited to $L/L_{\rm Edd} < 0.30$ \citep{McClintock+06}.  Because the NT model improves as the thickness and luminosity of the disc decrease (Table~5), {\it we conclude that use of the NT thin-disc model does not limit our accuracy.  Rather, it is the uncertainties in the input parameters $D$, $M$ and $i$ that strongly dominate the error in $a_*$.}


\section{Conclusions and Discussion}
\label{sec:conc}
The main conclusion of this paper is that observational errors in current measurements of black hole spin by the continuum-fitting method dominate over the errors incurred by using the idealized \citetalias{NovikovThorne73} model. We reached this conclusion by using 3D GRMHD simulations of thin discs to obtain realistic disc temperature profiles, then calculating the corresponding spectra, and finally fitting these spectra using the standard \textsc{XSPEC} model \textsc{KERRBB}.
For disc thicknesses \referee{$|h/r| \sim 0.04 - 0.08$}, the errors in $a_*$ are up to about 0.2, depending on the inclination, for a non-spinning black hole, and up to about 0.1, 0.03 and 0.01 for black holes with spins of 0.7, 0.9 and 0.98 respectively (\autoref{tab:spins-main}).
The errors in the spin estimates are particularly large at low spins and high inclinations, e.g., we find a spin estimate of \referee{0.15} for a non-spinning black hole viewed at an inclination angle of $75^\circ$. The results are \referee{quite close to} those obtained by \citet{ReynoldsFabian08} for the iron line fitting method \referee{(see Fig. 5 of their paper)}. Interestingly, we find that the fitted accretion rates are correct to within a few percent.

A new and important contribution in this paper is that we establish via the slim-disc and \textsc{BHSPEC} models an approximate correspondence between the disc thickness as calculated from GRMHD simulations (\autoref{sec:grmhd}) and the key disc observable $L/L_{\rm Edd}$ (\autoref{sec:slimdisk}). Even though the simulated discs considered in this paper are geometrically quite thin, \referee{$|h/r|\sim0.04-0.08$}, nevertheless it turns out that such discs correspond to fairly high luminosities, \referee{$L/L_{\rm Edd} \sim 0.4 - 0.7$}. For comparison, in observational work based on the continuum fitting method, the data-selection criterion $L/L_{\rm Edd}<0.3$ \citep{McClintock+06} is generally employed (which, from \autoref{fig:hr}, corresponds to $|h/r| \lesssim 0.03$). The validity and usefulness of this criterion can be best judged by examining the results for 411 observations of LMC X-3 in Steiner et al. (2010a, see their Figs. 2 and 3).  For $L/L_{\rm Edd}<0.3$, the inner disc radius $r_{\rm in}$ is very nearly constant, rising only slightly at luminosities above $L/L_{\rm Edd}\approx0.2$. However, $r_{\rm in}$ increases quite abruptly as the luminosity exceeds 30\% of Eddington. Remarkably, at these higher luminosities there is little scatter in the data and the increase in $r_{\rm in}$ is smooth and systematic.

\referee{It is difficult to say at this stage what the reason is for the above increase in the inner disc radius above the critical luminosity of $\sim 0.3L_{\rm Edd}$. Both the GRMHD and the slim-disc models predict that the inner disc radius should decrease \citep[see the discussion of the radiation edge in][]{Abramowicz+10}\footnote{\referee{The caveat, as \citet{Abramowicz+10} point out, is that there are various ways of defining the inner edge of the accretion disc, and for some definitions, the inner disc radius can increase when the luminosity increases beyond $\sim 0.3L_{\rm Edd}$ if the viscosity parameter is large enough ($\alpha \gtrsim 0.2$). However, the values of $\alpha$ that we see in our simulations are smaller, so this caveat does not present any problem.}}.
On the other hand, Li et al. (2010) were able to reproduce the observed increase by considering self-shadowing of the disc as a result of the off-midplane location of the disc photosphere.}
Interestingly, \citet{Abramowicz+10} find that the inner radius of the disc is fairly close to the \citetalias{NovikovThorne73} value for luminosities $L \lesssim 0.3 L_{\rm Edd}$, and that the inner edge decreases quite abruptly at higher values of $\dot M$. This, combined with the observed behaviour of $r_{\rm in}$ in LMC X-3, may be a hint that something qualitatively different happens at $L/L_{\rm Edd}\approx0.3$; perhaps energy advection or disc self-shadowing becomes suddenly more relevant.

One firm conclusion can be drawn from the results presented in this paper.  Since \autoref{tab:hr2} indicates that the modeling error decreases as the disc thickness decreases, whatever the behaviour of $r_{\rm in}$ may be above $L/L_{\rm Edd}\approx0.3$, at luminosities appropriate to the continuum-fitting method ($L<0.3L_{\rm Edd}$), where the disc will be geometrically very thin, the errors in the spin estimates will be even smaller than those quoted in \autoref{tab:spins-main}. Therefore, these errors are not a concern for the continuum-fitting method of measuring black hole spin.

We must note one caveat about comparing the density scale height from the GRMHD simulations with that from the slim-disc and \textsc{BHSPEC} models: the latter models do not include magnetic pressure. The increase in the photosphere height due to magnetic pressure could be as large as a factor of 2 \citep{Hirose+06}, although other studies have found more modest changes \citep{Blaes+06, Davis+09}.

\subsection*{Acknowledgments}
\referee{The authors wish to thank Jason Dexter for a critical reading of the manuscript, and the referee for several comments that helped improve the paper.} This work was supported by NASA grant NNX08AH32G and NSF grant AST-0805832. The simulations presented in this work were performed on the NASA cluster Pleiades and the TeraGrid cluster Kraken.


\bibliography{ms}

\begin{appendix}

\section{Matching a GRMHD model to the Page \& Thorne Solution}
\label{app:match}
\citet[hereafter \citetalias{PageThorne74}]{PageThorne74} define in their eqs (31a,b) two
quantities: (i) $f(r)$, which is proportional to the local disc flux
$F_{\rm com}(r)$ emitted from one side of the disc, as measured in the comoving frame of the fluid\footnote{The procedure for transforming the flux between the BL and comoving frames is described in \autoref{app:trans-flux}.}, and (ii) $w(r)$, which is proportional to the shear stress
$W_\phi^r$:
\begin{align}
f(r) &= 4\pi rF_{\rm com}/\dot{M}\label{eq:f(r)},\\
w(r) &= 2\pi rW_\phi^r/\dot{M}.
\end{align}
These quantities enter the angular momentum and energy
conservation laws via \citepalias[see][Eqs. 32a,b]{PageThorne74}
\begin{align}
(L^\dag-w)_{,r} &= fL^\dag, \label{eq:PTLconserve} \\
(E^\dag-\Omega w)_{,r} &= fE^\dag, \label{eq:PTEconserve}
\end{align}
where $L^\dag(r)$, $E^\dag(r)$ and $\Omega(r)$ are the specific
angular momentum, specific energy-at-infinity and the angular
velocity. \autoref{eq:PTLconserve} and \autoref{eq:PTEconserve}
lead to the further relation \citepalias[Eq. 33]{PageThorne74}
\begin{equation}
f=-\Omega_{,r} (E^\dag-\Omega L^\dag)^{-1} w.
\end{equation}

For a thin accretion disc, \citetalias{PageThorne74} show that $f$ has a general solution of
the form
\begin{equation}\label{eq:PTsolution}
f(r) = -\Omega_{,r} (E^\dag-\Omega L^\dag)^{-2}\left[
\int (E^\dag-\Omega L^\dag) L_{,r}^\dag dr + C\right],
\end{equation}
where $C$ is an integration constant.  They assume that the stress
vanishes at the ISCO, and thereby determine the value of $C$.  They
then obtain the following particular solution:
\begin{equation}
f_{\rm PT}(r) = -\Omega_{,r} (E^\dag-\Omega L^\dag)^{-2}
\int_{r_{\rm ISCO}}^r (E^\dag-\Omega L^\dag) L_{,r}^\dag dr.
\end{equation}
Eq. 15n in \citetalias{PageThorne74} gives an explicit analytical expression for $f_{\rm
PT}(r)$.

We are interested in the following more general problem.  We have a
GRMHD numerical solution of a thin disc that has reached inflow
equilibrium out to some radius $r_{\rm ie}$ (defined in \citealt{Penna+10}).  Beyond this radius,
however, we cannot trust the numerical results, so we would like to
match our simulation model to the \citetalias{PageThorne74} solution. This will allow
us to extrapolate the simulation \referee{beyond $r_{\rm ie}$ and even} beyond the radial range of the
numerical grid.  We wish to avoid the particular solution
$f_{\rm PT}$ given above since that assumes zero stress at the ISCO.
Instead, we fit for the value of the integration constant $C$ using
the simulation.  We also redefine the constant $C$ slightly so that
the fitting is made at $r_{\rm ie}$ rather than $r_{\rm ISCO}$.

Let us write \autoref{eq:PTsolution} as follows,
\begin{multline}\label{eq:mysolution}
f(r) = -\Omega_{,r} (E^\dag-\Omega L^\dag)^{-2}\left[
\int_{r_{\rm ie}}^r (E^\dag-\Omega L^\dag) L_{,r}^\dag dr + C\right], \\
\quad r\geq r_{\rm ie},
\end{multline}
where the new constant $C$ is to be determined from the simulation at $r=r_{\rm ie}$.
We can write
\begin{align}
\begin{split}
\int_{r_{\rm ie}}^r (E^\dag-\Omega L^\dag) L_{,r}^\dag dr &=
\int_{r_{\rm ISCO}}^r (E^\dag-\Omega L^\dag) L_{,r}^\dag dr \\
&\qquad - \int_{r_{\rm ISCO}}^{r_{\rm ie}} (E^\dag-\Omega L^\dag) L_{,r}^\dag dr
\end{split} \\
\begin{split}
&= \left[\frac{(E^\dag-\Omega L^\dag)^2}{-\Omega_{,r}} \right]_{r} f_{\rm PT}(r) \\
&\qquad - \left[\frac{(E^\dag-\Omega L^\dag)^2}{-\Omega_{,r}} \right]_{r_{\rm ie}} f_{\rm PT}(r_{\rm ie}).
\end{split}
\end{align}
Substituting in \autoref{eq:mysolution}, we obtain the result we seek:
\begin{multline}\label{eq:result}
f(r)=f_{\rm PT}(r)-\frac{[(E^\dag-\Omega L^\dag)^2/\Omega_{,r}]_{r_{\rm ie}}}
{[(E^\dag-\Omega L^\dag)^2/\Omega_{,r}]_{r}} f_{\rm PT}(r_{\rm ie}) \\
- \frac{\Omega_{,r}}{(E^\dag-\Omega L^\dag)^2}C, \quad r \geq r_{\rm ie},
\end{multline}
where
\begin{equation}
C=-\left[\frac{(E^\dag-\Omega L^\dag)^2}{\Omega_{,r}}\right]_{r_{\rm
      ie}} 4\pi r_{\rm ie}F_{\rm com}(r_{\rm ie})/\dot{M}.
\end{equation}
Except for $F_{\rm com}(r_{\rm ie})/\dot{M}$, the local disc flux of the
simulation at $r=r_{\rm ie}$, all the other quantities are obtained
from the idealized \citetalias{PageThorne74} model and can be evaluated at any $r$ outside
the matching radius $r_{\rm ie}$.  With this matching procedure, the
local disc flux from the converged region of the simulation can be
extrapolated to arbitrary radii.

The choice of the matching radius $r_{\rm ie}$ is an important issue. We would like to use as large a radius as possible, while still staying within the inflow equlibrium radius, but we need to ensure that the choice of the matching radius does not affect the luminosity profiles $d (L/\dot M) / d(\ln r)$ significantly. \referee{\autoref{fig:rtrans} shows the luminosity profiles for $a_*=0$, 0.7, 0.9 and 0.98 (bottom to top). For each spin, the cluster of solid lines shows the luminosity profiles for $r_{\rm ie} / r_{\rm ISCO} =$ 1.1, 1.2, 1.3, 1.4, and the dashed line shows the \citetalias{NovikovThorne73} profile. It is clear that for all the models except $a_*=0.98$, the choice of the matching radius has little effect on the luminosity profiles. This turns out to be particularly true for $r_{\rm ie} / r_{\rm ISCO} < 1.3$, as a closer look reveals.} We therefore use $r_{\rm ie} = 1.3 r_{\rm ISCO}$ in those three models. \referee{The $a_*=0.98$ model, however, did not attain inflow equilibrium out to a sufficiently large radius. Its luminosity profiles therefore show more sensitivity to the matching radius. Hence we choose $r_{\rm ie} = 1.2 r_{\rm ISCO}$ for this model. This is a conservative choice, since it results in a larger deviation from the \citetalias{NovikovThorne73} luminosity.
The predicted errors in the spin estimates shown for this model in \autoref{tab:spins-main} are therefore likely to be overestimates.}

\fignar{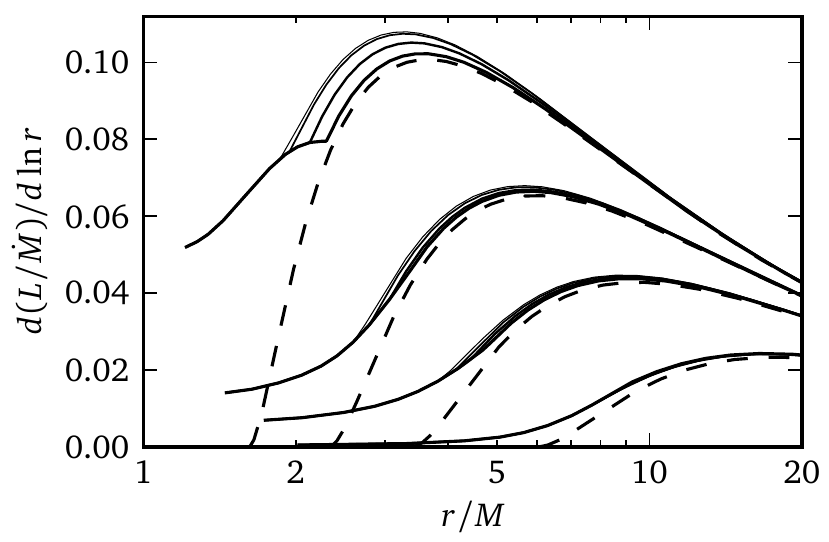}{fig:rtrans}{\referee{Similar figure to \autoref{fig:lum}, except that the solid line for each spin in \autoref{fig:lum} is replaced by a cluster of solid lines showing the luminosity profiles for different values of the matching radius, $r_{\rm ie} / r_{\rm ISCO} =$ 1.1, 1.2, 1.3, 1.4 (solid lines, top to bottom).}}

\section{Transformation Between the Boyer-Lindquist and Comoving Frames}
\label{app:trans}

\subsection{Transformation of vectors and one-forms}
\label{app:trans-vec}
To transform the four-momentum, we need to find the orthonormal basis of the comoving frame, which we do using the Gram-Schmidt orthonormalization procedure as described in \citet{BeckwithHawleyKrolik08} and \citet{ShcherbakovHuang11}. Let us denote the comoving-frame basis vectors by $e_{(\nu)}$
and the fluid 4-velocity by $u^\mu$.
In the comoving frame, since the fluid is at rest, its four-velocity $u$ is equal to $e_{(t)}$. This is a coordinate-independent statement; it is true in any frame. We can thus denote the components of the $e_{(\nu)}$ in the Boyer-Lindquist (BL) frame by
\begin{equation}
\begin{split}
   e^\mu_{(t)} &= u^{\mu} = (u^t, u^r, u^\theta, u^\phi), \\
   e^\mu_{(r)} &= (\lambda_3, \lambda_4, 0, \lambda_5), \\
   e^\mu_{(\theta)} &= (\lambda_6, \lambda_7, \lambda_8, \lambda_9), \\
   e^\mu_{(\phi)} &= (\lambda_1, 0, 0, \lambda_2),
\end{split}
\end{equation}
where the $\lambda_i$ are determined by imposing orthonormality, i.e., that $e^\mu_{(\nu)}e_\mu^{(\psi)} = \delta^{(\psi)}_{(\nu)}$. (The index $\mu$ in $e^\mu_{(\nu)}$ is lowered using the BL metric $g_{\alpha\beta}$, while the index $(\nu)$ is raised using the Minkowski metric $\eta^{(\alpha)(\beta)} = \text{diag}(-1, 1, 1, 1)$, since we want the comoving frame to be locally flat as well.) The $e^\mu_{(\nu)}$ are the components of the transformation matrix for four-vector components from the comoving frame into the BL frame. Some algebra gives
\begin{gather}
\begin{split}
   e^\mu_{(t)} \,\,\, &=  \,\,\, (u^t, \,\,\, u^r, \,\,\, u^\theta, \,\,\, u^\phi), \\
   e^\mu_{(r)} \,\,\, &= \,\,\, (u_r u^t, \,\,\, -(u_t u^t + u_\phi u^\phi), \,\,\, 0, \,\,\, u_r u^\phi) \,\,\, / \,\,\, N_r, \\
   e^\mu_{(\theta)} \,\,\, &= \,\,\, (u_\theta u^t, \,\,\, u_\theta u^r, \,\,\, 1 + u_\theta u^\theta, \,\,\, u_\theta u^\phi) \,\,\, / \,\,\, N_\theta, \\
   e^\mu_{(\phi)} \,\,\, &= \,\,\, (u_\phi, \,\,\, 0, \,\,\, 0, \,\,\, -u_t) \,\,\, / \,\,\, N_\phi, \\
\intertext{and its inverse}
   e^{(t)}_\mu \,\,\, &= \,\,\, (-u_t, \,\,\, -u_r, \,\,\, -u_\theta, \,\,\, -u_\phi), \\
   e^{(r)}_\mu \,\,\, &= \,\,\, (u_r u_t, \,\,\, -g_{rr}(u_t u^t + u_\phi u^\phi), \,\,\, 0, \,\,\, u_r u_\phi) \,\,\, / \,\,\, N_r, \\
   e^{(\theta)}_\mu \,\,\, &= \,\,\, (u_\theta u_t, \,\,\, u_\theta u_r, \,\,\, g_{\theta\theta}(1 + u_\theta u^\theta), \,\,\, u_\theta u_\phi) \,\,\, / \,\,\, N_\theta, \\
   e^{(\phi)}_\mu \,\,\, &= \,\,\, -\Delta \sin^2\theta \,\,\, (u^\phi, \,\,\, 0, \,\,\, 0, \,\,\, -u^t) \,\,\, / \,\,\, N_\phi, \\
\intertext{with}
   N_r^2 &= -g_{rr} (u_t u^t + u_\phi u^\phi) (1 + u_\theta u^\theta) \\
   N_\theta^2 &= g_{\theta\theta} (1 + u_\theta u^\theta) \\
   N_\phi^2 &= -(u_t u^t + u_\phi u^\phi) \Delta \sin^2\theta, \\
   \Delta &= r^2 + a^2 - 2Mr.
\end{split}
\end{gather}
Here, $a$ is the (dimensionful) black hole spin: $a \equiv a_* GM/c^2.$ The transformation laws for vectors and one-forms are then given by
\begin{gather}
   X^\mu = e^\mu_{(\nu)} X^{(\nu)}, \qquad X^{(\mu)} = e^{(\mu)}_\nu X^\nu, \notag\\
   X_\mu = e^{(\nu)}_\mu X_{(\nu)}, \qquad X_{(\mu)} = e^\nu_{(\mu)} X_\nu.
\end{gather}

\subsection{Transformation of the flux}
\label{app:trans-flux}
To transform the radiation flux from the BL frame (which is what the GRMHD simulations give) into the comoving frame of the fluid, we use the following method. The comoving flux is defined as
\begin{equation}
F_{\rm com} = \frac{dE_{\rm com}}{dA_{\rm com} dt_{\rm com}} \equiv \frac{dE_{\rm com}}{d^3V_{\rm com}},
\end{equation}
where $d^3V_{\rm com}$ is the 3-volume in the comoving frame. We first relate the energy in the fluid frame to that in the BL frame.
Let $e_{\tilde\Omega}$ be the energy emitted per unit solid angle in the comoving frame. A ray of light emitted into a solid angle $d\tilde\Omega$ in a direction $(\tilde\theta, \tilde\phi)$ (where $\tilde\theta$ is measured with respect to the normal to the disc, i.e., the $e_{(\theta)}$ direction, and the $\tilde\phi=0$ direction is arbitrary) will then have an energy-momentum four-vector given by
\begin{align}
   dp^{(\mu)} &= e_{\tilde\Omega} d\tilde\Omega (1, \sin\tilde\theta \sin\tilde\phi, \cos\tilde\theta, \sin\tilde\theta \cos\tilde\phi), \\
\intertext{or, lowering the index, and remembering that the metric in the comoving frame is $\eta_{(\alpha)(\beta)} = \text{diag}(-1, 1, 1, 1)$,}
   dp_{(\mu)} &= e_{\tilde\Omega} d\tilde\Omega (-1, \sin\tilde\theta \sin\tilde\phi, \cos\tilde\theta, \sin\tilde\theta \cos\tilde\phi). \\
\intertext{Transforming this into the BL frame, we get}
   dp_\mu &= e^{(\nu)}_\mu dp_{(\nu)}. \\
\intertext{The energy-at-infinity of that ray as measured in the BL frame then is}
   dE &= -dp_t = -e_t^{(\nu)} dp_{(\nu)}. \\
\intertext{The total energy emitted in all directions is}
   E &= \int dE = -e_t^{(\nu)} \int dp_{(\nu)} \\
   &= -e_t^{(\nu)} \int d\tilde\Omega e_{\tilde\Omega} (-1, \sin\tilde\theta \sin\tilde\phi, \cos\tilde\theta, \sin\tilde\theta \cos\tilde\phi).
\intertext{The emission profiles we are interested in are isotropic ($e_{\tilde\Omega}$ = constant) or limb-darkened ($e_{\tilde\Omega} \propto 2 + 3\cos\tilde\theta$). For both of these, the second and fourth terms in the brackets vanish. The third term is multiplied by $e_t^{(\theta)} \propto u_\theta = 0$ for the thin discs that we consider here, for which the fluid velocity is purely in the equatorial plane. So we are left with}
   E &= -u_t \int d\tilde\Omega e_{\tilde\Omega} = -u_t E_{\rm com}.
\end{align}

Next, we need to transform the 3-volume $d^3V_{\rm com}$. Consider first a four-volume in the BL frame bounded by $dt$, $dr$, $d\theta$ and $d\phi$. The proper four-volume $\sqrt{-g} dt dr d\theta d\phi = (\sqrt{-g_{tr\phi}} dt dr d\phi) (\sqrt{g_{\theta\theta}} d\theta)$ is an invariant. (Here, $g$ is the determinant of the covariant BL metric, and $g_{tr\phi}$ is the determinant of the $t-r-\phi$ part of the metric.) Also, since the transformation into the comoving frame of the fluid involves a boost perpendicular to the $\theta$-direction, the proper length in the $\theta$-direction (which is the term in the second set of brackets in the last expression) is also invariant. (This can be checked explicitly by examining the transformation of the vector $X^\mu = (0, 0, d\theta, 0)$ into the comoving frame.) Therefore, the proper three-volume $\sqrt{-g_{tr\phi}} dt dr d\phi$ is also invariant. We therefore have
\begin{equation}
d^3V_{\rm com} = \sqrt{-g_{tr\phi}} dt dr d\phi = r dr d\phi dt \text{ in the equatorial plane}.
\end{equation}
The comoving flux then becomes
\begin{align}
F_{\rm com} &= \frac{dE}{(-u_t)r dr d\phi dt} \\
        &= \frac{F}{-u_t},
\end{align}
which is the required transformation for the flux.

\section{Ray-Tracing Grid}
\label{app:grid}
The grid in the image plane is generated using plane polar coordinates $(b, \beta)$. The radial grid points form a geometric series, $b_m = b_0 q^m$, $m = 0, \ldots, N_b$, where $q$ is a number slightly larger than unity. This gives us high resolution close to the center of the image plane, which is necessary for resolving the inner region of the accretion disc. The spacing in the angular direction is linear: $\beta_n = \beta_0 + n \delta\beta$, $n = 0, \ldots, N_\beta$. In addition, the grid is ``squeezed'' along the direction defined by the projection of the black hole spin axis onto the image plane (the ``y''-axis), by an amount proportional to $\cos i$, to account for the inclination of the observer. Therefore, the radial grid becomes a set of concentric ellipses; the Cartesian coordinates of the grid vertices are $x_{mn} = b_m \cos\beta_n$, $y_{mn} = b_m \sin\beta_n \cos i$. We choose $b_0 = 1$, which is inside the black hole shadow for all spins and inclinations, i.e., it is small enough that any rays with $b<b_0$
fall into the black hole and can therefore be ignored. We also choose $b_{N_b} = 10000$, which is large enough to correctly produce the low-energy end of the spectrum in the energy range of interest (0.1 keV to 10 keV).

The flux in \autoref{eq:fobsdef} then becomes
\begin{align}
   F_{\rm obs} &= \oneby{D^2} \int I_\nu \, dx \, dy \\
           &= \frac{\cos i}{D^2} \int I_\nu \, b \, db \, d\beta \\
           &= \frac{\cos i}{D^2} \int I_\nu \, b^2 \, d(\log b) \, d\beta.
\end{align}
Since the values of $\log b$ and $\beta$ are equally spaced, we can now use a higher order integration method like the Simpson's 1/3-rd rule to perform the integration \citep[see, e.g.,][]{AbramowitzStegun72}. This gives considerably higher accuracy and faster convergence than a na\"ive approach that simply sums up the product of $I_\nu$ in each grid cell with the area of that cell.

\end{appendix}

\end{document}